\documentstyle[prb,aps,multicol,epsf]{revtex}
\tighten
\narrowtext
\newcommand \be {\begin{equation}}
\newcommand \ee {\end{equation}}
\newcommand \bea {\begin{eqnarray}}
\newcommand \eea {\end{eqnarray}}
\begin{document}
\draft

\title{Magnetotransport in 2D lateral superlattices with smooth
disorder: Quasiclassical theory of commensurability oscillations} 
\author{A.~D.~Mirlin$^{a,b,*}$, E.~Tsitsishvili$^{b,c}$, 
and P.~W\"olfle$^{a,b}$}
\address{
$^a$ Institut f\"ur Nanotechnologie, Forschungszentrum Karlsruhe,
D-76021 Karlsruhe, Germany\\
$^b$ Institut f\"ur Theorie der Kondensierten Materie,
Universit\"at Karlsruhe, D-76128 Karlsruhe, Germany\\
$^c$ Institute of Cybernetics, Euli 5, 380086 Tbilisi, Georgia}
\date{\today}
\maketitle
\begin{abstract}
Commensurability oscillations in the magnetoresistivity of a
two-dimensional electron gas in a two-dimensional lateral superlattice
are studied in the framework of quasiclassical transport
theory. It is assumed that the impurity scattering is of small-angle
nature characteristic for currently fabricated high-mobility
heterostructures. The shape of the modulation-induced
magnetoresistivity $\Delta\rho_{xx}$ depends on
the value of the parameter $\gamma\equiv \eta^2 ql/4$, where $\eta$
and $q$ are the strength and the wave vector of the modulation, and
$l$ is the transport mean free path. For $\gamma\ll 1$, the
oscillations are described, in the regime of not too strong magnetic
fields $B$, by perturbation theory in $\eta$ as applied earlier to the
case of one-dimensional modulation. At stronger fields, where 
$\Delta\rho_{xx}$ becomes much larger than the Drude
resistivity, the transport takes the advection-diffusion form
(Rayleigh-B\'enard convection cell) with a large P\'eclet number,
implying a much slower ($\propto B^{3/4}$) increase of the oscillation
amplitude with $B$.  If $\gamma\gg 1$, the transport at low $B$ is
dominated by the modulation-induced chaos (rather than by
disorder). The magnetoresistivity drops exponentially and the
commensurability oscillations start to develop at the magnetic
fields where the motion takes the form of the adiabatic drift. Conditions
of applicability, the role of the type of disorder, and
the feasibility of experimental observation are discussed.

\end{abstract}

\begin{multicols}{2}

\section{Introduction}
\label{s1}

Electronic transport in semiconductor nanostructures is one of 
central issues of research in  modern condensed matter
physics, see {\it e.g.} \cite{beenakker91,ferry} for reviews. In
particular, transport properties of a two-dimensional electron gas
(2DEG) subject to a periodic potential (lateral superlattice) with a
period much shorter than the electron transport mean free path (but
much larger than the Fermi wave length) have
been intensively studied during the last decade. In a pioneering
experiment \cite{weiss} Weiss
{\it et al.} discovered that a weak one-dimensional (1D) modulation 
with wave vector ${\bf q}\parallel {\bf e}_x$ induces strong
commensurability oscillations of the magnetoresistivity
$\rho_{xx}(B)$ (while showing almost no effect on $\rho_{yy}(B)$ and
$\rho_{xy}(B)$), with the minima satisfying the
condition $2R_c/a=n-1/4$, $n=1,2,\ldots$, where $R_c$ is the
cyclotron radius and $a=2\pi/q$ the modulation wave
length. The quasiclassical nature of these commensurability oscillations
was demonstrated by Beenakker \cite{beenakker}, who showed that the
interplay of the cyclotron motion and the superlattice potential
induces a drift of the guiding center along $y$ axis, with an
amplitude squared oscillating as
$\cos^2(qR_c-\pi/4)$ (this is also reproduced by a quantum-mechanical
calculation, see \cite{theory-qm}). While describing nicely the period
and the phase of the experimentally observed oscillations, the result
of \cite{beenakker}, however, failed to explain the observed rapid
decay of the oscillation 
amplitude with decreasing magnetic field. The cause for this
discrepancy was in the treatment of disorder: while 
Ref.~\onlinecite{beenakker} assumed isotropic impurity scattering, in
experimentally relevant high-mobility semiconductor heterostructures
the random potential is 
very smooth and induces predominantly small-angle scattering, with
the total relaxation rate $\tau_s^{-1}$ much exceeding the momentum
relaxation rate $\tau^{-1}$. The theory of commensurability
oscillations in the situation of smooth disorder was worked out by two
of us in \cite{mw98}. It was found that the small-angle scattering  
drastically modifies the dependence of the oscillation amplitude on the
magnetic field $B$, leading to a much stronger damping of the oscillations
with decreasing $B$, in full agreement with experimental data (see
also Monte-Carlo simulations in \cite{boggild} and numerical solution
of the Boltzmann equation in \cite{menne98}). The small-angle nature of 
the scattering plays also an important role in the theoretical description
\cite{mtw1} of the low-field magnetoresistance dominated by channeled 
orbits \cite{beton90}. Summarizing, one can say that the theory
of magnetoresistivity in 1D lateral superlattices is by now fairly
well understood and provides a quantitative description of the experiment. 

In contrast, the situation with two-dimensional (2D) superlattices is
by far less clear. Experimental studies of transport in samples with
weak 2D modulation (which can be obtained by superimposing two 1D
modulations with equal amplitudes and orthogonal wave vectors) 
\cite{alves89,gerhardts91,liu91,lorke91,weiss92,
steffens98,albrecht99,chowdhury00} have demonstrated that the
commensurability oscillations can be observed in this case as well,
but their amplitude is to some extent suppressed as compared to the
case of 1D modulation. However, no clear answer concerning the
conditions for this suppression, as well as its magnitude, can
be drawn from the literature. 

The present state of the theory is equally controversial. It was
stated in \cite{gerhardts92} that within the quasiclassical approach a
contribution to the resistivity tensor induced by a 2D modulation,
\be
\label{e1}
V(x,y)=\eta E_F(\cos qx + \cos qy)\ , \qquad \eta\ll 1,
\ee
where $E_F$ is the Fermi energy,
is given by a trivial generalization of the result of \cite{beenakker}
(for 1D modulation), yielding 
\be
\label{e2}
\Delta\rho_{xx}^{\rm 2D}=\Delta\rho_{yy}^{\rm 2D}=
\Delta\rho_{xx}^{\rm 1D}({\bf q}\parallel {\bf e}_x)\ .
\ee
Experimental deviations from the results of this perturbative-in-$V$
approach were attributed to inadequacy of the
quasiclassical treatment.
In complete contrast, the authors of Ref.~\onlinecite{grant00} claimed
that the perturbation theory of \cite{beenakker} fails in 2D. They
argued, in particular, that
since in a symmetric ($V_x=V_y$) 2D modulation equipotential contours
(along which the guiding center drifts) are closed, the
modulation-induced correction to resistivity should vanish.  

The aim of this paper is to present a systematic theoretical analysis
of the quasiclassical transport in a weak 2D superlattice in the
presence of a smooth random potential. We will
show that the overall picture is much richer than any of the two
above-mentioned extremes (proposals of \cite{gerhardts92} and
of \cite{grant00}) would suggest. We will also demonstrate that, as in
the case of a 1D modulation, the 
nature of the disorder crucially affects the magnetoresistivity. 

It is worth mentioning that in Ref.~\onlinecite{pfannkuche92} a purely
quantum-mechanical interpretation of experimentally observed
commensurability oscillations was proposed, related to the miniband
structure of Landau levels induced by modulation. The authors of
\cite{pfannkuche92} argued that this structure suppresses the
quasiclassical effect of the modulation. We note, however, that for
typical experimental parameters the miniband splitting is small compared
to the total scattering rate $1/\tau_s$, implying that the disorder
broadening washes out the miniband structure \cite{footnote1}. In this
situation, the quasiclassical theory should provide an adequate
description of the effect. Indeed, the large amplitude of commensurability
oscillations (of the same order as in samples with 1D modulation)
observed in many experiments
\cite{alves89,lorke91,steffens98,albrecht99,chowdhury00} strongly
suggests their classical origin. 
Only recently, first experimental observations of indications of the
miniband structures (smeared considerably by disorder) in the regime
of strong Shubnikov-de Haas oscillations have been reported
\cite{butterfly}.  In any case, the quasiclassical theory
of the commensurability oscillations should serve as a reference point
also for analysis of the quantum effects. 

\section{Advection-diffusion transport in a 2D periodic potential}
\label{s2}

In the absence of disorder and in a sufficiently strong magnetic field
$B$ (the condition will be discussed in Sec.~\ref{s3} below) the
motion of a particle 
in the periodic potential (\ref{e1}) is a drift along the
equipotential contours $V_{\rm eff}(x,y)={\rm const}$. Here $x$ and
$y$ are the coordinates of the guiding center, and $V_{\rm eff}$ is
obtained by averaging (\ref{e1}) along the cyclotron orbit, 
\be
\label{e3}
V_{\rm eff}(x,y)=\eta E_F J_0(qR_c)(\cos qx +\cos qy)\ ,
\ee
where $R_c=v_F/\omega_c$ is the cyclotron radius, $v_F$ the Fermi
velocity, and $\omega_c=eB/mc$ the cyclotron frequency. 
The drift velocity ${\bf v}_d(x,y)$ is given by
\be
\label{e4}
{\bf v}_d(x,y)={1\over eB^2}{\bf \nabla}V_{\rm eff}(x,y)\times{\bf B}.
\ee
Switching to the rotated coordinates $X=(x+y)/\sqrt{2}$, 
$Y=(y-x)/\sqrt{2}$, we find the
components of the drift velocity
\bea
v_{dX} & = & - v \sin(\pi Y /d)\cos (\pi X/d)\ , \label{e5}\\
v_{dY} & = &   v \sin(\pi X /d)\cos (\pi Y/d)\ , \label{e6}
\eea
where $d=a/\sqrt{2}$ and
\bea
\label{e7}
v & = & {1\over\sqrt{2}} \eta v_F q R_c J_0(qR_c) \nonumber \\
  & \simeq & 
{1\over\sqrt{\pi}} \eta v_F (q R_c)^{1/2}\cos(qR_c-\pi/4)
\eea
(in the last line we assumed $qR_c\gg 1$). Eqs.~(\ref{e5}), (\ref{e6})
determine an incompressible flow $v_{dX}=\partial\Psi/\partial Y$, 
$v_{dY}=-\partial\Psi/\partial X$, with a stream function
\be
\label{e8}
\Psi(X,Y)={vd\over\pi}\cos(\pi X/d)\cos(\pi Y/d)\ .
\ee
The streamlines of this flow are closed and belong to one of the 
cells $n-1/2<X/d<n+1/2$, $m-1/2<Y/d<m+1/2$, with integer $m$ and
$n$, which form a square lattice. The streamlines $X=d(n+1/2)$ and
$Y=d(m+1/2)$ separating the cells are called separatrices.

We turn now to the effect of disorder. The small-angle scattering
leads to diffusion of the guiding center with the diffusion constant
\be
\label{e9}
D={v_F^2\tau\over 2(\omega_c\tau)^2}
\ee
(we assume $\omega_c\tau\gg 1$). The limits of applicability of this
diffusion approximation will be discussed below, see Sec.~\ref{s4}. We
are thus left with 
a transport problem of the advection-diffusion type. This kind of
problem has been studied in the context of hydrodynamics since long
ago, see \cite{isichenko} for review. More specifically, with the
periodic stream function (\ref{e8}) we are facing the problem of the
diffusive transport in a Rayleigh-B\'enard convection cell
\cite{isichenko,moffat,shraiman}. The nature
of the transport depends crucially on the value of the P\'eclet number
\be
\label{e10}
P={vd\over D} \sim {\eta ql\over (qR_c)^{3/2}}\ ,
\ee
where $l=v_F\tau$ is the transport mean free path. 
If $P\ll 1$, the impurity scattering dominates over the advection. In
this case the particle ``does not realize'' that the equipotential
contours are closed, and the correction to the conductivity tensor due
to the periodic potential can be calculated using perturbative
expansion in $P$. Therefore, in this regime, the
perturbative expansion in the modulation strength $\eta$
is valid, yielding \cite{mw98,footnote3} 
\begin{equation}
{\Delta\rho_{xx}\over\rho_0}  =  {\eta^2 ql \over 4} Q
{\pi\over\sinh\pi\mu} J_{i\mu}(Q)J_{-i\mu}(Q)\ ,
\label{e11}
\end{equation}
where $\rho_0=h/e^2 k_Fl$ is the Drude resistivity ($k_F$ is the Fermi
wave vector),
\begin{equation}
\mu={Q\over qv_F\tau_s}
\left[1-\left(1+{\tau_s\over\tau}Q^2\right)^{-1/2}\right]\ ,
\label{e12}
\end{equation}
and we introduced the dimensionless parameter $Q=qR_c$ convenient to
characterize the strength of the magnetic field. At low 
magnetic fields, $Q\gg Q_{\rm dis}$, with $Q_{\rm dis}=(2ql/\pi)^{1/3}$, the
oscillations are exponentially damped by disorder, and the
magnetoresistivity saturates at the value
\begin{equation} 
{\Delta\rho_{xx}\over\rho_0}  =  {\eta^2 ql\over 4} \equiv \gamma 
\label{e13}
\end{equation}
(we will use below the parameter $\gamma$ introduced here in order to
classify different transport regimes). Note that at still lower
magnetic fields, $Q>Q_{\rm ch}$, with $Q_{\rm ch}=2/\eta$, and for a sufficiently
strong modulation, $\eta^{3/2} ql\gg 1$, an additional strong
magnetoresistivity occurs, dominated by the channeled orbits
\cite{beton90,mtw1}. We will not consider this region of magnetic
fields in the present
paper. In strong fields, $Q\ll Q_{\rm dis}$, the amplitude of oscillations
increases as $B^3$,
\begin{equation}
{\Delta\rho_{xx}\over\rho_0}={(\eta ql)^2\over \pi Q^3}\cos^2(Q-\pi/4)\ .
\label{e14}
\end{equation}
Note that the condition of validity of Eq.~(\ref{e14}), $P\ll 1$, is
equivalent to $\Delta\rho_{xx}/\rho_0\ll 1$. 

\begin{figure}
\narrowtext
{\epsfxsize=7cm\centerline{\epsfbox{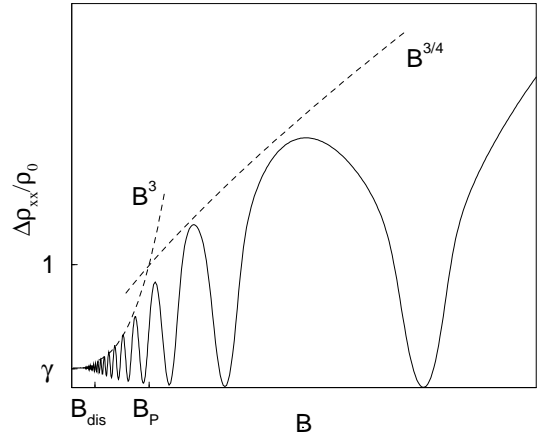}}}
\vskip0.3cm
\caption{Schematic representation of the magnetoresistivity
$\Delta\rho_{xx}(B)$ induced by the modulation in the case $\gamma\ll
1$. Characteristic points $B_{\rm dis}$ and $B_P$ on the magnetic
field axis (corresponding to $Q=Q_{\rm dis}$ and $Q=Q_P$, see the text)
are shown. Below $B_{\rm dis}$ the oscillations are exponentially
damped and the magnetoresistivity saturates at
$\Delta\rho_{xx}=\gamma$, while at $B=B_{\rm P}$ the $B^3$-behavior of
$\Delta\rho_{xx}$ changes to a much slower, $B^{3/4}$-increase.}
\label{fig1}
\end{figure}

In the opposite limit of large P\'eclet number, $P\gg 1$, the
transport is determined by a narrow boundary layer around the square
network of separatrices (``stochastic web''). The width $d_b$ of this
layer 
can be estimated from the condition that the particle can diffuse
through the layer in a period of the advection motion, yielding
$d_b\sim(Dd/v)^{1/2}$. This leads to an effective diffusion
coefficient \cite{moffat}
\be
\label{e15}
D_{\rm eff}\sim d_b v \sim (Dvd)^{1/2}\sim P^{1/2}D\ .
\ee
Calculation of the numerical coefficient in (\ref{e15}) requires a
more involved analysis \cite{shraiman}, the result being 
\be
\label{e16}
D_{\rm eff}=C(Dvd)^{1/2}\ , \qquad C\simeq 0.62.
\ee
Substituting (\ref{e7}), (\ref{e9}) in (\ref{e16}), we find the
modulation-induced magnetoresistivity
\be
\label{e17}
{\Delta\rho_{xx}\over\rho_0}=(8\pi)^{1/4}C(\eta ql)^{1/2}Q^{-3/4}
|\cos(Q-\pi/4)|^{1/2}\ .
\ee
Therefore, in sufficiently strong magnetic fields, where the relative 
correction to the Drude resistivity becomes large, 
$\Delta\rho_{xx}/\rho_0\gg 1$, its $B$-dependence changes to 
$\Delta\rho_{xx}/\rho_0\propto B^{3/4}$. The two formulas (\ref{e14})
and (\ref{e17}) match at $Q=Q_{\rm P}\equiv [0.13(\eta ql)^2]^{1/3}$, their
conditions of validity being $Q\gg Q_{\rm P}$ and $Q\ll Q_{\rm P}$,
respectively. This behavior of the magnetoresistivity is illustrated
in Fig.~\ref{fig1}.

In the following two sections we will analyze conditions of validity of
the two approximation used in the above derivation: the drift
approximation for the motion in the periodic potential and the
diffusion approximation for the impurity scattering. We will also
briefly discuss the transport regimes which take place when one of
these approximations is violated.

\section{Role of adiabaticity of the motion in the periodic potential:
drift vs chaotic diffusion}
\label{s3}

The drift approximation reflects the adiabatic nature of the electron
motion in the periodic potential and is applicable provided the shift
$\delta$ of the guiding center after one cyclotron revolution is small
compared to the modulation period $a$. Noting that
$\delta_X=(2\pi/\omega_c)v_{dX}$ and that according to Eq.~(\ref{e7})
the average squared drift velocity is equal to
\be
\label{e18}
\langle v_{dX}^2\rangle={\eta^2\over 4\pi}v_F^2
Q\cos^2(Q-\pi/4)\ ,
\ee
we find for the typical shift $\delta=\langle \delta_X^2\rangle^{1/2}$,
\be
\label{e19}
{\delta\over a} = {\eta\over (4\pi)^{1/2}}Q^{3/2}
|\cos(Q-\pi/4)|\ .
\ee
Therefore, the adiabatic condition $\delta/a\ll 1$ can be rewritten as
$Q\ll Q_{\rm ad}$, where $Q_{\rm ad}=(4\pi/\eta^2)^{1/3}$. Comparing $Q_{\rm ad}$ with the
value $Q_{\rm dis}$ determining the point where the drift motion breaks down
due to the impurity scattering (see above), we find
\be
\label{e20}
{Q_{\rm dis}\over Q_{\rm ad}}=\left({\eta^2 ql\over 2\pi^2}\right)^{1/3}\ .
\ee
We see that this ratio is determined by the parameter $\gamma$ defined
in Eq.~(\ref{e13}), which governs also the relation between $Q_{\rm dis}$ and
$Q_{\rm P}$, 
\be
\label{e21}
{Q_{\rm P}\over Q_{\rm dis}}\simeq (0.2 \eta^2 ql)^{1/3}\ .
\ee
Therefore, one should analyze separately two regimes, $\gamma\ll 1$
and $\gamma\gg 1$.

In the case $\gamma\ll 1$ we have $Q_{\rm ad}\gg Q_{\rm dis}\gg Q_{\rm P}$. The first
inequality means that the effect of the randomization of the guiding
center position by the periodic potential is weaker than the analogous
effect of the random potential. Therefore, the former effect can be
neglected, and the point $Q=Q_{\rm ad}$ is irrelevant for the behavior of
magnetoresistivity. We also note that $Q_{\rm ch}\gg Q_{\rm ad}$, so that the
region of the low-field magnetoresistivity (channeled orbits),
$Q\gtrsim Q_{\rm ch}$ is parametrically separated from the region of
commensurability oscillations, $Q\lesssim {\rm min}\{Q_{\rm dis},Q_{\rm ad}\}$.

The assumed condition $\gamma\ll 1$ is typically fulfilled in
an experiment if the modulation strength  $\eta$ is of order
of a few percent. For example, taking typical experimental values of
the parameters, $ql=1000$, $\eta=0.03$, we get $\gamma\simeq 0.22$ and
$Q_{\rm P}\simeq 4.9$, $Q_{\rm dis}\simeq 8.6$, $Q_{\rm ad}\simeq 24$, $Q_{\rm ch}\simeq 67$, so
that the above inequalities $Q_{\rm P}\ll Q_{\rm dis}\ll Q_{\rm ad}\ll Q_{\rm ch}$ are indeed
reasonably satisfied. 
 
We turn now to the consideration of the opposite limit $\gamma\gg 1$,
which is realized for sufficiently strong modulation. Now we have
$Q_{\rm ad}\ll Q_{\rm dis}\ll Q_{\rm P}$, implying that $Q_{\rm dis}$
and $Q_{\rm P}$ become 
irrelevant. In the whole region $Q\gtrsim Q_{\rm ad}$ the non-adiabatic
electron dynamics leads to chaotic diffusion \cite{geisel}.
The transport in this regime is similar to that in the
region $Q\gg Q_{\rm dis}$ for $\gamma\ll 1$, with disorder replaced by the
modulation-induced chaos. Therefore, the magnetoresistivity is given
in this regime by Eq.~(\ref{e13}) (note that $\Delta\rho_{xx}$ in this
formula does not depend explicitly on disorder), with the oscillations
being exponentially damped. 

\begin{figure}
\narrowtext
{\epsfxsize=7cm\centerline{\epsfbox{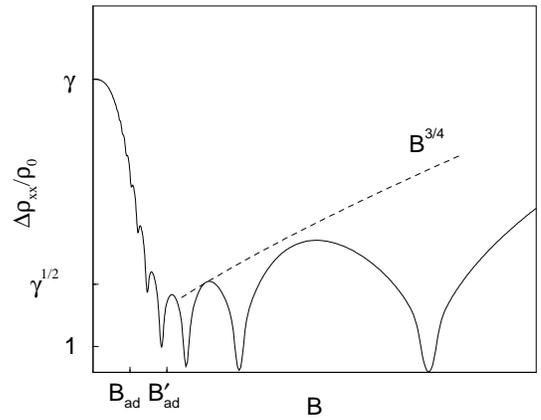}}}
\vskip0.3cm
\caption{Schematic representation of the  magnetoresistivity
$\Delta\rho_{xx}(B)$ in the case $\gamma\gg 1$. 
The magnetoresistivity starts to drop exponentially and the
commensurability oscillation appear at the value $B_{\rm ad}$ of the
magnetic field where the motion in the periodic potential takes the
form of an adiabatic drift. At $B\sim B'_{\rm ad}$ the disorder
starts to dominate over the non-adiabatic effects, leading to a 
$B^{3/4}$-increase of the oscillation amplitude.}
\label{fig2}
\end{figure}

With magnetic field  increasing beyond the point $Q=Q_{\rm ad}$ the system
enters the adiabatic regime, with the advection-diffusion type of
transport, as considered in Sec.~\ref{s2}. One can thus expect
Eq.~(\ref{e17}) to be applicable. It is clear, however, that
Eq.~(\ref{e17}) does not match the formula (\ref{e13})
[valid for $Q\gg Q_{\rm ad}$] at $Q\sim Q_{\rm ad}$. Let us
demonstrate that there will be an 
additional, logarithmically narrow, intermediate regime between the
regions of validity of these two formulas. The reason is that even at
$Q\ll Q_{\rm ad}$ the picture of adiabatic drift in a periodic potential
(\ref{e1}) is only approximate, though its violation is exponentially
weak. This leads, in the absence of any disorder, to the formation of a
stochastic boundary layer of a width $\tilde{d_b}$ falling off
exponentially with magnetic field \cite{fogler}
\bea
\label{e22}
{{\tilde d_b}\over d} & \propto & 
\exp\left(-{\omega_c d\over 2 v}\right) \nonumber\\
& = & \exp\left[-{\pi\over 2\sqrt{2}|\cos(Q-\pi/4)|}
\left({Q_{\rm ad}\over Q}\right)^{3/2}\right]
\eea
(we omit preexponential factors here). The underlying physics is
analogous to that discussed in Sec.~\ref{s2} in the case of a large
P\'eclet number, with the only difference that the diffusion is now
not due to disorder but rather due to violation of adiabaticity. The
width ${\tilde d_b}$ exceeds the disorder-induced one, $d_b$, in a
narrow range $Q_{\rm ad}\gg Q \gg Q'_{\rm ad}$, where 
$Q'_{\rm ad}\sim Q_{\rm ad}(\ln\gamma)^{-2/3}$. Within this
logarithmically narrow 
interval of magnetic fields, the transport is still fully determined
by the modulation, with the macroscopic diffusion coefficient 
$\tilde{D}_{\rm eff} \sim v \tilde{d_b}$, and the
magnetoresistivity drops off exponentially (Fig.~\ref{fig2}),
\be
\label{e23}
{\Delta\rho_{xx}\over\rho_0}\propto 
\exp\left[-{\pi\over 2\sqrt{2}|\cos(Q-\pi/4)|}
\left({Q_{\rm ad}\over Q}\right)^{3/2}\right]\ ,
\ee
from $\Delta\rho_{xx}/\rho_0\sim\gamma$ at $Q\sim Q_{\rm ad}$ to 
$\Delta\rho_{xx}/\rho_0\sim\gamma^{1/2}$ at $Q\sim Q'_{\rm ad}$, with
a simultaneous rapid development of commensurability oscillations.  
At $Q<Q'_{\rm ad}$ the disorder-induced diffusion becomes more
efficient than 
that due to violation of adiabaticity of the motion in the periodic
potential, so that Eq.~(\ref{e17}) becomes applicable, and the
magnetoresistivity starts to increase with magnetic field as
$B^{3/4}$, as illustrated in Fig.~\ref{fig2}. 

\section{Role of the range of disorder}
\label{s4}

In this section we discuss the condition of validity of the diffusion
approximation used in Sec.~\ref{s2}. This approximation corresponds
formally to the limit of zero time interval between two successive
scattering events. In a real situation this interval is, however,
finite, its average value being equal to $\tau_s$. Correspondingly,
the r.m.s. shift of the guiding center in a scattering event is given
by
\bea
\label{e24}
\delta_s &\equiv& \langle(\delta x)^2\rangle^{1/2}
= \langle(\delta y)^2\rangle^{1/2} \nonumber\\ 
&=& (2D\tau_s)^{1/2}=  R_c(\tau_s/\tau)^{1/2}\ .
\eea
The ratio $\tau_s/\tau\ll 1$ is of the order of $1/(k_F\xi)^2$, where
$\xi$ is the correlation length of the random potential set by the
width of the undoped spacer between the 2DEG and the impurity layer,
since an electron is typically scattered by a small angle
$\delta\phi\sim 1/k_F\xi$. In order for the diffusion-advection
description of the transport presented in Sec.~\ref{s2} to be
justified, the following two conditions should be fulfilled: (i)
$\delta_s\ll a$ and (ii) $\tau_s\ll a/v$, which can be equivalently
rewritten as
\be
\label{e25}
Q\ll 2\pi(\tau/\tau_s)^{1/2} \equiv Q_{\rm diff}^{(1)}
\ee
and
\be
\label{e26}
Q\cos^2(Q-\pi/4)\ll 8\pi^3 \left({\tau\over \tau_s}\right)^2
{1\over (\eta ql)^2} \equiv Q_{\rm diff}^{(2)},
\ee
respectively. Let us estimate typical values of $Q_{\rm diff}^{(1)}$
and $Q_{\rm diff}^{(2)}$. Using the same parameters as above
($ql=1000$, $\eta=0.03$) and the ratio $\tau/\tau_s=50$
(characteristic for the best of currently fabricated high-mobility
heterostructures \cite{smet98}), we find $Q_{\rm diff}^{(1)}\simeq 45$ 
and $Q_{\rm diff}^{(2)}\simeq 690$, so that the diffusion
approximation is justified in the whole range of interest, $Q\lesssim
Q_{\rm ad}$, where the oscillations are observed. We see, however,
that this conclusion relies heavily on the small-angle nature of the
scattering, $\tau/\tau_s\gg 1$. If we would assume in the above
example isotropic scattering, $\tau/\tau_s=1$, we would get 
$Q_{\rm diff}^{(1)}\simeq 6.3$ 
and $Q_{\rm diff}^{(2)}\simeq 0.28$, so that the diffusion
approximation would not be valid in the whole region of oscillations. 
We analyze now what happens if at least one of the conditions
(\ref{e25}), (\ref{e26}) is violated. The following three situations
can be distinguished:

(A) Eq.~(\ref{e25}) violated, Eq.~(\ref{e26}) fulfilled. In this case
the picture of the transport is as follows: the guiding center drifts
during a time $\sim\tau_s$ along an equipotential contour to a small
distance $\sim v\tau_s\ll a$, then makes a large jump of a range $\sim
\delta_s\gg a$, etc. The disorder-induced correction to resistivity is
thus equal to
\be
\label{e27}
{\Delta\rho_{xx}\over\rho_0}= 
{2\langle v_{dx}^2\rangle\tau_s^2\over\delta_s^2} =
{v^2\tau\tau_s\over 2 R_c^2} \ll 1\ ,
\ee
with $v$ given by Eq.~(\ref{e7}). This result can be obtained by
a perturbative expansion of the type used in \cite{beenakker,mw98}
($\Delta\rho_{xx}/\rho_0\propto\eta^2$), which is determined by the
fact that a drifting particle does not close a contour before it is
scattered far away. The correction (\ref{e27}) shows according to
(\ref{e7}) the conventional
commensurability oscillations $\propto\cos^2(Q-\pi/4)$, but it is
small, $\Delta\rho_{xx}/\rho_0\ll 1$, since it can be presented in the
form
$$
{\Delta\rho_{xx}\over\rho_0}\sim \left({v\tau_s\over a}\right)^2
\left({a\over\delta_s}\right)^2\ ,
$$
both factors being much less than unity.

(B) Both Eqs.~(\ref{e25}) and (\ref{e26}) violated. In this case the
particle typically completes the drift contour before it is scattered
to a large distance $\delta_s\gg a$. The resistivity correction can
thus be estimated as
\be
\label{e28}
{\Delta\rho_{xx}\over\rho_0}\sim {a^2\over\delta_s^2}
={a^2\over R_c^2}{\tau\over\tau_s}\ll 1\ .
\ee
Since (\ref{e28}) does not depend on the drift velocity, the
magnetoresistivity does not show oscillations as a function of
$Q$. More precisely, there will be a remnant of oscillations in the
form of narrow dips in the vicinity of zeros of the drift velocity,
$Q=\pi(n-1/4)$, where Eq.~(\ref{e26}) will be fulfilled and the
resistivity will be given by Eq.~(\ref{e27}).

(C) Eqs.~(\ref{e25}) fulfilled, Eq.~(\ref{e26}) violated. The
transport in this regime is determined by a boundary layer of the
width $\delta_s$ near separatrices, so that
$$
D_{\rm eff}\sim {a\delta_s\over\tau_s}\ ,
$$
and, correspondingly,
\be
\label{e29}
{\rho_{xx}\over\rho_0} = {D_{\rm eff}\over D} \sim 
{a(\tau/\tau_s)^{1/2}\over R_c}\gg 1\ .
\ee
As in the case (B), Eq.~(\ref{e29}) does not depend on $v$, so that
the oscillations are washed out, except for narrow vicinities of the
points $Q=\pi(n-1/4)$, where the condition (\ref{e26}) will be
restored and thus the result (\ref{e17}) [or, still closer to a zero
of the drift velocity, Eq.~(\ref{e14})] will be applicable. 

Summarizing the results for the regimes (B) and (C), one can conclude
that if the condition $Q\lesssim Q_{\rm diff}^{(2)}$ is violated, the
modulation-induced resistivity is strongly suppressed (as compared to
what the diffusion approximation would predict) and the
commensurability oscillations are washed out. This explains, in
particular, why essentially no oscillations were observed in the
numerical simulations of Ref.~\onlinecite{grant00} for a symmetric
square superlattice (\ref{e1}). These authors assumed a model of
isotropic scattering, $\tau_s=\tau$, which typically leads to the
violation of both inequalities 
(\ref{e25}) and (\ref{e26}), as shown above. The resulting
magnetoresistivity is weak, Eq.~(\ref{e28}), and shows only remnants of
oscillations, in agreement with the simulations in \cite{grant00}. 

\section{Summary and discussion}
\label{s5}

In this article we have presented a systematic theoretical analysis of
the quasiclassical magnetotransport of 2D electrons in a 2D lateral
superlattice. The amplitude of the modulation-induced
magnetoresistivity, and in particular of the commensurability
oscillations, depends crucially on the nature of disorder (see
Sec.~\ref{s4}). We assumed small-angle impurity scattering induced
by a smooth random potential characteristic for high-mobility
heterostructures. In the region of existence of commensurability
oscillations the transport is determined by the interplay of the
advection of the guiding center in the periodic potential and the
diffusion due to impurity scattering. The shape of the
magnetoresistivity depends on the dimensionless parameter $\gamma$,
Eq.~(\ref{e13}). 

For small $\gamma$ (corresponding typically to a
modulation strength not exceeding a few percent) the
magnetoresistivity is given by the perturbative formulas (\ref{e11})
-- (\ref{e14}) up to the point $Q\sim Q_{\rm P}$, where the correction
$\Delta\rho_{xx}$ becomes of the order of the Drude resistivity
$\rho_0$. For higher magnetic fields the P\'eclet number $P$
characterizing the advection-diffusion problem becomes large and the
transport is determined by a narrow boundary layer around a square
network of separatrices. As a result, the $B^3$-dependence of the
oscillation amplitude characteristic for the perturbative 
($Q>Q_{\rm P}$) regime crosses over to a much slower
$B^{3/4}$-increase at $Q<Q_{\rm P}$, see Eq.~(\ref{e17}). 

For $\gamma\gg 1$ (which is typically valid for the modulation
strength $\eta$ larger than $10 \div 15 \%$) the oscillations are
damped at low magnetic fields not by disorder (as in the perturbative
regime) but by the modulation-induced chaotic diffusion. The
oscillations become observable at $Q\sim  Q_{\rm ad}$ where the
motion of electrons in the superlattice potential acquires the form of
adiabatic drift. Since the violation of adiabaticity is exponentially
small, the magnetoresistivity drops exponentially in a logarithmically
narrow interval of magnetic fields, 
$Q_{\rm ad}\ln^{-2/3}\gamma< Q<Q_{\rm ad}$, Eq.~(\ref{e23}). At higher
magnetic fields the impurity scattering starts to dominate over the
non-adiabatic processes and thus to determine the diffusion constant of
the advection-diffusion problem, so that the commensurability
oscillations take the same form (\ref{e17}) as in the
large-$P$ limit of the $\gamma\ll 1$ regime. 

We have demonstrated that the transport regimes discussed in
Sec.~\ref{s2} and \ref{s3}, {\it i.e.} the perturbative regime
(advection-diffusion with small $P$), the stochastic web regime
(advection-diffusion with $P\gg 1$), and the chaos-dominated regime
with an exponential fall-off of the magnetoresistivity at the
adiabaticity threshold,  are within the range of typical experimental
parameters. Experimental identification of these regimes and a
quantitative analysis of the experimental data on the basis of the
theory presented here is thus not only highly desirable but also
practically feasible. 

Finally, let us note that in this paper, as well as in the earlier
study of 1D modulation \cite{mw98}, we treated disorder as a
collision term in the Boltzmann equation, thus neglecting memory
effects. On the other hand, it has been shown recently that for smooth
disorder these effects may become important with increasing magnetic
field \cite{magres}, eventually leading to an adiabatic character of the
motion in a random potential \cite{fogler} or a random magnetic field
\cite{rmf}, and consequently to an exponential drop of
magnetoresistivity (in the absence of any modulation). Though, to our
knowledge, this behavior has not yet been observed experimentally in
the low-field magnetotransport of a 2D electron gas, the progress in
fabrication of better samples with larger spacers may make it
experimentally relevant in the near future. Also, a positive
magnetoresistance due to memory effects, followed by the
adiabatic fall-off of the resistivity is observed in magnetotransport of
composite fermions near half-filling of the lowest Landau level, where
the effective disorder has the form of a smooth random magnetic
field \cite{rmf}. Understanding of the effect of a 1D or 2D periodic
modulation in the situation of such a non-Boltzmann transport in the
unmodulated system would require a separate theoretical analysis.

\section*{Acknowledgments}

This work was supported by the SFB195 der Deutschen
Forschungsgemeinschaft, the DFG-Schwerpunktprogram
``Quanten-Hall-Systeme'', the German-Israeli Foundation, and the INTAS 
grant 97-1342.

\end{multicols}


\begin{references}
\bibitem{footnote1} In
Ref.~\onlinecite{pfannkuche92} the small-angle nature of scattering leading
to $1/\tau_s\gg 1/\tau$ was neglected and unrealistically small values
of the collision broadening were assumed. Specifically, the Landau
level broadening $\Gamma_0=0.015\,{\rm meV}\times\sqrt{B(\rm T)}$ used
in calculations of \cite{pfannkuche92} corresponds to $\tau_s\simeq
800\,{\rm ps}$, which exceeds by more than two orders of magnitude the
values $\tau_s\sim 5\,{\rm ps}$ characteristic for currently
fabricated high-mobility samples.
\bibitem{footnote3} There is a misprint in the corresponding formula
[Eq.~(50)] of Ref.~\onlinecite{mw98}: the factor $QS$ should be
replaced by $Q^2S$.
\end{references}
\end{document}